\documentclass[prl,reprint,
preprintnumbers]{revtex4-1}
\pdfoutput=1
\usepackage{hyperref}
\usepackage[T1]{fontenc}

\usepackage{mathrsfs}
\usepackage{enumitem}

\usepackage{amsmath,amssymb,amsfonts,amsxtra,mathrsfs,graphics,graphicx,amsthm,epsfig,bm,longtable,float,color,tikz,mathtools,xfrac,footnote}
\restylefloat{table}
\usepackage{dsfont}
\usepackage{lipsum}
\usepackage{textgreek}
\usetikzlibrary{decorations.pathmorphing}
\usetikzlibrary{decorations.markings}
\usetikzlibrary{quotes,arrows.meta}
\usetikzlibrary{arrows,decorations.markings,calc,fadings,decorations.pathreplacing,patterns,decorations.pathmorphing,positioning}
\usepackage{tikz-cd}

\newcommand{\cf}{\textit{cf.}}

\newcommand{\ie}{\textit{i.e.}}

\newcommand{\ii}{\mathrm{i}}
\newcommand{\?}{\;\!}

\newcommand{\be}{\begin{equation}} \newcommand{\ee}{\end{equation}}
\newcommand{\bea}{\begin{equation} \begin{aligned}} \newcommand{\eea}{\end{aligned} \end{equation}}

\def\U{\mathrm{U}}

\newcommand{\rd}{\mathrm{d}}

\newcommand{\wb}{\overline}
\newcommand{\wt}{\widetilde}

\DeclareMathOperator{\re}{\mathbb{R}e}
\DeclareMathOperator{\im}{\mathbb{I}m}

\newcommand{\pd}{\partial}

\newcommand{\cE}{\mathcal{E}}

\newcommand{\cH}{\mathcal{H}}
\newcommand{\cI}{\mathcal{I}}

\newcommand{\cN}{\mathcal{N}}

\newcommand{\cX}{\mathcal{X}}

\newcommand{\bC}{\mathbb{C}}

\begin{document}

\title{Gravitational Blocks: Symplectic Covariance Unveiled}

\date{\today}

\author{Seyed Morteza Hosseini}
\affiliation{Department of Physics, Imperial College London, London, SW7 2AZ, UK}
\email{s.hosseini@imperial.ac.uk}

\begin{abstract}

By working in a symplectically covariant real formulation of special K\"ahler geometry, we propose and give strong evidence for a canonical BPS partition function for AdS$_2 \times_w M_2$ near--horizon geometries with arbitrary rotation and generic magnetic and electric charges. Here, $M_2$ is either a two--sphere or a spindle. We also show that how the attractor equations and the Bekenstein--Hawking entropy can be obtained from an extremization principle.

\end{abstract}

\pacs{}
\keywords{}


\maketitle

{\it Introduction.}---String theory offers a valuable theoretical testing ground for exploring the quantum properties of black holes,
thus helping us to address questions that would otherwise be insurmountable.
In particular, Strominger and Vafa in their seminal work \cite{Strominger:1996sh} demonstrated the viability of string theory
by providing a microscopic interpretation for the Bekenstein--Hawking entropy of certain classes of BPS black holes in asymptotically Minkowski spacetimes. More recently, starting with \cite{Benini:2015eyy},
progress made towards understanding the entropy of asymptotically anti de Sitter (AdS) black holes microscopically (see \cite{Zaffaroni:2019dhb} for a recent review).
In this picture, black holes are thermodynamic ensembles of microstates in a holographic dual conformal field theory (CFT).

For a BPS black hole with angular momentum $J$ and a set of conserved magnetic and electric charges $( P^I , Q_I )$ in $\cN = 2$ matter--coupled supergravity,
the Bekenstein--Hawking area law formula $S_{\text{BH}} = \frac{A}{4}$ \footnote{We use units where the Newton’s constant is fixed as $G_{\text{N}} = 1$.},
where $A$ is the area of the event horizon,
may be encoded in an entropy functional that reduces the problem of computing the entropy into an extremization principle.
This is equivalent to the attractor mechanism \cite{Ferrara:1996dd, Ferrara:1995ih,Ooguri:2004zv}:
at the horizon the scalar fields approach an attractor point, the critical points of the entropy functional, associated to the black hole charges.
The extremization principle identifies the entropy as a Legendre transform of a \emph{complex} functional $\cE ( P^I, \varphi^I , \epsilon )$
with respect to the electric and angular potentials $(\varphi^I , \epsilon)$,
and this suggests a thermodynamic interpretation of the attractor mechanism.
Note, that with respect to the magnetic charges $P^I$ one is therefore dealing with a microcanonical ensemble!

For AdS black holes in string and M--theory, a unifying entropy functional based on gluing \emph{gravitational blocks} was proposed in \cite{Hosseini:2019iad} (see also \cite{Faedo:2021nub, Faedo:2022rqx, Boido:2022iye}).
Remarkably, the `mixed free energy' $\cE ( P^I, \varphi^I , \epsilon )$ obtained in this way from the thermodynamic entropy matches precisely
the logarithm of the partition function of the holographic dual field theory in the large $N$ limit, when the latter is known.
For other cases it provides a prediction for the logarithm of the microstate degeneracy up to terms that vanish in the large $N$ limit.
In the approach of \cite{Hosseini:2019iad} the magnetic and electric charges are \emph{not} treated on equal footing, and thus there is \emph{no} manifest symplectic covariance;
a central feature of $\cN = 2$ theories based on vector multiplets (see for example \cite{LopesCardoso:2019swk}).
One question that naturally comes to mind is: is there a way to maintain symplectic covariance?
This forms the main subject of this letter.

Consider AdS$_2 \times_w M_2$ near--horizon geometries with arbitrary rotation and generic magnetic and electric charges.
When $M_2$ is either a two--sphere or a spindle (a two--sphere with conical singularities at the poles), we conjecture that in the limit of large charges
$$
 Z_{\text{can.}} ( \Phi , \epsilon ) \approx e^{\pi \ii \frac{\cH ( \Phi )}{\epsilon}} \, ,
$$
where $Z_{\text{can.}} ( \Phi , \epsilon )$ is a \emph{fully canonical} BPS partition function
depending on the symplectic vector of electric and magnetic potentials $\Phi = \{ \varphi^I ; \chi_I \}$,
and $\cH ( \Phi )$ is the \emph{Hesse potential} \cite{Hitchin:1999hmi} of the scalar manifold, that is a symplectic function.
Remarkably, $\frac{\cH ( \Phi )}{\epsilon}$ is related to the free energy $\cE ( P^I, \varphi^I, \epsilon )$ via a Legendre transform,
which replaces $P^I$ by $\chi_I$ as independent variables.
Then, the black hole entropy is obtained by extremizing the entropy functional
$$
 \cI ( \Phi , \epsilon ) = \log Z_{\text{can.}} ( \Phi , \epsilon ) - \pi \ii \left( \varphi^I Q_I - \chi_I P^I - \ii \epsilon J \right) ,
$$
with respect to the chemical potentials $(\Phi , \epsilon )$, subject to a constraint that depends on the model,
and evaluating it at its critical points.
We show that the extremal values of $( \Phi, \epsilon )$
are completely determined by the values of the symplectic vector of scalars at the North Pole and the South Pole of $M_2$ in the near--horizon region,
thus providing an explicit realization of the attractor mechanism.

We note that a symplectically covariant entropy functional based on the Hesse potential has been put
forward for static BPS black holes in \emph{ungauged} four--dimensional $\cN = 2$ supergravity earlier \cite{LopesCardoso:2006ugz}.
See \cite{LopesCardoso:2019swk} for a review and other related developments.
We also notice strong similarities between our formalism and those based on Sen's entropy functional \cite{Sen:2005wa,Sen:2005iz}.

{\it Gravity's attractive blocks.}---To set the scene, we will review the extremization principle of \cite{Hosseini:2019iad} that captures the Bekenstein--Hawking entropy
of AdS$_2 \times S^2_\epsilon$ near--horizon geometries, with magnetic and electric charges encoded in the symplectic vector $\Gamma = \{P^I ; Q_I \}$ and angular momentum $J$
\footnote{The magnetic charges are subject to a linear constraint (as imposed by the BPS equations)
and the remaining conserved quantities obey some (non--)linear regularity constraints
that decrease the number of independent conserved charges for regular BPS black holes.},
in the four--dimensional gauged $\cN = 2$ supergravity coupled to $n_\text{V} = 3$ abelian vector multiplets
\footnote{We will focus on the $stu$ model where $n_{\text{V}} = 3$, but the specific examples suggest that
our method can be generalized to $\cN = 2$ supergravity models coupled to $n_{\text{V}}$ vector multiplets and $n_{\text{H}}$ hypermultiplets.}.
The Lagrangian of the latter is completely determined by a single holomorphic function of the scalars, $F( X^I )$, which is called the prepotential,
and the symplectic vector of gauging parameters $G = \{g^I ; g_I \}$.
First, we define the gravitational block
\be
 \label{GR:blocks:def}
 B \big( X^I _{(\sigma)}, \epsilon_{(\sigma)} \big) \equiv \frac{1}{2 \ii} \frac{F ( X^I_{(\sigma)} )}{\epsilon_{(\sigma)}} \, ,
\ee
that depends on the scalars $X^I_{(\sigma)}$, $I = 0, \ldots, n_{\text{V}}$,
evaluated at the North Pole $(\sigma = + 1)$ or the South Pole $(\sigma = - 1)$ of $S^2_\epsilon$,
and the chemical potential $\epsilon$ for the angular momentum on $S^2_\epsilon$.
Observe that the gravitational block \eqref{GR:blocks:def} is \emph{not} a symplectic function, but transforms in a rather complicated way.
Let us \emph{formally} decompose $X^I_{(\sigma)}$ into real and imaginary parts,
$X^I_{(\sigma)} = \varphi^I_{(\sigma)} + \ii \epsilon_{(\sigma)} P^I$ with $\varphi^I_{(\sigma)}  \in \bC$ and $\epsilon_{(\sigma)} \in \bC$,
and introduce the functional $\cE_{A/id} ( P^I, \varphi^I , \epsilon )$
that is obtained by gluing gravitational blocks \eqref{GR:blocks:def}
\footnote{The value of the R--symmetry magnetic flux distinguishes between two types of supersymmetry preserving backgrounds \cite{Hristov:2011ye,Hristov:2011qr}.}:
\begin{enumerate}[label=(\roman*)]
 \item
 \emph{$A$--gluing} (supersymmetry on the $S^2_\epsilon$ is realized with a topological $A$--twist):
  \bea
   \label{hhz:A-gluing}
   \cE_{A} ( P^I, \varphi^I , \epsilon ) \equiv \frac{1}{2 \ii}
   \left[ \frac{F \big( \varphi^I + \ii \epsilon P^I \big)}{\epsilon}
   + \frac{F \big( \varphi^I - \ii \epsilon P^I \big)}{- \epsilon} \right] .
  \eea
 \item
  \emph{Identity gluing} (otherwise): 
  \bea
   \label{hhz:id-gluing}
   \cE_{id} ( P^I, \varphi^I , \epsilon ) \equiv \frac{1}{2 \ii}
   \left[ \frac{F \big( \varphi^I + \ii \epsilon P^I \big)}{\epsilon}
  + \frac{F^* \big( - \varphi^I - \ii \epsilon P^I \big)}{- \epsilon} \right] .
 \eea
 Here, $F^* ( X^I )$ denotes the \emph{formal conjugate} of the prepotential $F ( X^I )$ that is obtained by replacing any explicit factor of $\ii$ with $- \ii$.
\end{enumerate}
The functional $\cE_{A/id} ( P^I, \varphi^I , \epsilon )$ is defined within a \emph{mixed ensemble},
where the electric charges $Q_I$ and the angular momentum $J$ fluctuate, while the magnetic charges $P^I$ are kept fixed.
Thus, the independent variables are the chemical potentials $(\varphi^I, \epsilon)$ for the electric charges and rotation, and the magnetic charges $P^I$.
The Bekenstein--Hawking entropy is then obtained by extremizing the entropy functional
\bea
 \label{HHZ:entropy:function}
 \cI ( P^I, \varphi^I , \epsilon ) & \equiv \pi \ii \left( \cE_{A/id} ( P^I , \varphi^I , \epsilon ) - \varphi^I Q_I + \ii \epsilon J \right) \\
 & - \lambda \left( g_I \varphi^I - \alpha \epsilon - 1 \right) ,
\eea
with respect to the chemical potentials $(\varphi^I , \epsilon)$,
\be
 \begin{aligned}
 \label{mixed:attractor}
 Q_I + \lambda g_I & = \frac{\pd \cE_{A/id}}{\pd \varphi^I} \, , \quad I = 1, \ldots, n_{\text{V}} \, , \\
 J - \ii \alpha \lambda & = \frac{\pd \cE_{A/id}}{\pd \epsilon} \, ,
 \end{aligned}
\ee
and evaluating it at its critical points, $\cI ( P^I , \varphi^I , \epsilon ) |_{\eqref{mixed:attractor}} \equiv S_{\text{BH}} ( P^I , Q_I , J )$.
Here, we introduced the Lagrange multiplier $\lambda \in \bC$ to impose the constraint
\be
 \label{const:phi:epsilon}
 g_I \varphi^I - \alpha \epsilon - 1 = 0 \, ,
\ee
where $\alpha$ is a model dependent constant.
This constraint among chemical potentials is required by supersymmetry and the existence of a smooth black hole horizon.
Observe, that $e^{\pi \ii \cE_{A/id} ( P^I, \varphi^I , \epsilon )}$ has the natural interpretation of the black hole partition function, in the limit of large charges, in a mixed thermodynamic ensemble,
$$
 Z_{\text{mixed}} ( P^I , \varphi^I , \epsilon ) = \sum_{Q_I, J} d_{\text{micro}} ( P^I , Q_I , J ) e^{\pi \ii ( \varphi^I Q_I - \ii \epsilon J ) } \, ,
$$
in which magnetic charges $P^I$ are fixed integers, while electric charges $Q_I$ and the angular momentum $J$ are summed over weighted by the chemical potentials as $e^{\pi \ii ( \varphi^I Q_I - \ii \epsilon J )}$.
Here, the number $d_{\text{micro}}$ of black hole microstates with charges $( P^I , Q_I , J)$ is related to the mixed partition function through a Laplace transform with respect to $(\varphi^I , \epsilon)$,
\begin{widetext}
 $$
  d_{\text{micro}}  ( P^I , Q_I , J ) = \oint \rd \varphi \? \rd \epsilon \?
  \delta \left( g_I \varphi^I - \alpha \epsilon - 1 \right) 
  Z_{\text{mixed}} ( P^I, \varphi^I , \epsilon ) \?
  e^{- \pi \ii \left( \varphi^I Q_I - \ii \epsilon J \right)} \, ,
 $$
\end{widetext}
where $\delta (x)$ is the Dirac delta function, $\rd \varphi \equiv \prod_{I} \rd \varphi^I$,
and the chemical potentials $(\varphi^I , \epsilon)$ are taken to be complex and integrated along a contour encircling the origin.
Thus, in the limit of large charges, \cf\;\eqref{HHZ:entropy:function},
\be
 \label{Z:mixed}
 \log Z_{\text{mixed}} ( P^I, \varphi^I , \epsilon ) \approx \pi \ii \cE_{A/id} ( P^I, \varphi^I , \epsilon ) \, ,
\ee
that is interpreted as the free energy of the black hole in the mixed ensemble.
Here, $\approx$ means asymptotic equality in the limit of large charges.

The mixed ensemble is natural in the framework of equivariant localization as strongly suggested by \eqref{hhz:A-gluing} and \eqref{hhz:id-gluing}.
However, it has the disadvantage that the independent variables $(P^I, \varphi^I)$ do not constitute a symplectic vector, which obscures symplectic covariance.

{\it Hesse block and attractor mechanism.}---In this section we propose a unifying entropy functional that maintains manifest symplectic covariance.
We consider AdS$_2 \times_w M_2$ near--horizon geometries with magnetic and electric charges $(P^I , Q_I )$ and angular momentum $J$,
where $M_2$ is either a two--dimensional sphere $S^2_\epsilon$ or a spindle $\Sigma$ (a two--sphere with conical singularities at the poles).

Let us assume that the special K\"ahler manifold parameterized by the scalar fields $X^I$ is a symmetric space,
and introduce the quartic form $I_4$ that is invariant under symplectic transformations.
Define the \emph{Hesse block}
\be
 \label{Hesse:def}
 H ( \Phi , \epsilon ) \equiv \frac{2}{\epsilon} \cH ( \Phi ) \, ,
\ee
where $\cH ( \Phi ) = \sqrt{I_4 ( \Phi )}$ is the \emph{Hesse potential} of the conical affine special K\"ahler manifold associated with the scalar moduli space \cite{Hitchin:1999hmi}.
The Hesse potential is homogeneous of degree two with respect to the electric and magnetic potentials $\Phi = \{ \varphi^I ; \chi_I \}$ (that are taken to be complex),
and in the real formulation of special K\"ahler geometry, it plays a role similar to that of the holomorphic prepotential $F (X^I)$.
The Hesse block is twice the Legendre transform of the $\cE_{A/id} ( P^I, \varphi^I, \epsilon )$ functional with respect to the magnetic charges $P^I$,
thereby replacing $P^I$ by $\chi_I$ as independent variables
\footnote{When $J = 0$, the Hesse potential $\cH ( \Phi )$ is related to the prepotential $F ( X^I )$ by a Legendre transformation that replaces $\im ( X^I )$ with $\re ( F_I )$ as an independent field \cite{Cortes:2001bta}.
Here, $F_I \equiv \pd F ( X^I ) / \pd X^I$ are the holomorphic periods.},
\be
 \label{H:Leg:F}
 H ( \varphi^I , \chi_I , \epsilon ) = 2 \left( \cE_{A/id} ( P^I, \varphi^I , \epsilon ) -\chi_I P^I \right) ,
\ee
where $\chi_I = {\pd \cE_{A/id}}/{\pd P^I}$.
The latter expresses $P^I$ as a function of $(\varphi^I , \chi_I, \epsilon)$, locally,
and inserting this expression in the right hand side of \eqref{H:Leg:F} yields \eqref{Hesse:def}
\footnote{$\chi_I = {\pd \cE_{A/id}}/{\pd P^I}$ in general gives complex values for $P^I$.
The real part of this expression reproduces the physical charges.}.

The entropy functional \eqref{HHZ:entropy:function} is now replaced by
\bea
 \label{Me:entropy:function}
 \cI ( \Phi , \epsilon ) & \equiv \pi \ii \left( \frac{1}{2} H ( \Phi , \epsilon ) - \varphi^I Q_I + \chi _I P^I + \ii J \epsilon \right) \\
 & - \lambda \left( g_I \varphi^I - g^I \chi_I - \alpha \epsilon - 1 \right) .
\eea
Indeed, extremization of \eqref{Me:entropy:function} with respect to $( \Phi , \epsilon )$ yields
\be
 \label{PQ:dH=0}
 P^I = \frac{\ii}{\pi} g^I \lambda - \frac12 \frac{\pd H}{\pd \chi_I} \, , \qquad
 Q_I = \frac{\ii}{\pi} g_I \lambda + \frac12 \frac{\pd H}{\pd \varphi^I} \, ,
\ee
for $I = 1, \ldots, n_{\text{V}}$, and
\be
 \label{J:dH=0}
 J = \frac{\alpha}{\pi} \lambda + \frac{\ii}{2} \frac{\pd H}{ \pd \epsilon} \, .
\ee
Substituting \eqref{PQ:dH=0} and \eqref{J:dH=0} into \eqref{Me:entropy:function},
and noting that the Hesse block $H (\Phi , \epsilon)$ is homogeneous of degree one,
we conjecture that
\be
 \label{I:lambda:dH=0}
 \cI ( \Phi , \epsilon ) \Big|_{\eqref{PQ:dH=0} \text{ and } \eqref{J:dH=0}} = \lambda ( P^I , Q_I , J ) 
 \equiv S_{\text{BH}} ( P^I , Q_I , J )  \, .
\ee
The requirement that \eqref{I:lambda:dH=0} be real positive, as it should be for a physical black hole solution, then fixes
\be
 \begin{aligned}
 \label{PQJ:attractor}
 P^I & = - \frac12 \re \left( \frac{\pd H}{\pd \chi_I} \right) , \qquad
 Q_I = \frac12 \re \left( \frac{\pd H}{\pd \varphi^I} \right) , \\
 J & = - \frac{1}{2} \im \left( \frac{\pd H}{ \pd \epsilon} \right) ,
 \end{aligned}
\ee
that are just the attractor equations written in terms of the quantities $( \varphi^I , \chi_I , \epsilon )$.
Attractor phenomenon implies that at the horizon the scalars $X^I_{\text{NP/SP}}$ take values
that only depend on the black hole charges $(P^I, Q_I , J)$ and not on their asymptotically specified values.
We shall see that the extremal values of the chemical potentials $( \varphi^I , \chi_I , \epsilon )$
are completely determined by the values of the symplectic vector of scalars $\cX = \{ X^I ; F_I (X^I ) \}$,
where $F_I \equiv \pd F ( X^I ) / \pd X^I$ are the holomorphic periods, at the North Pole and the South Pole of $M_2$ in the near--horizon region.

To make the connection with microstate degeneracies $d_{\text{micro}}  ( P^I , Q_I , J )$,
let us define the black hole partition function in the \emph{canonical ensemble}
$$
 Z_{\text{can.}} ( \varphi^I , \chi_I , \epsilon ) = \sum_{P^I, Q_I, J} d_{\text{micro}} ( P^I , Q_I , J )
 e^{\pi \ii ( \varphi^I Q_I - \chi_I P^I - \ii \epsilon J ) } \, ,
$$
which is invariant under the various duality symmetries,
only if the chemical potentials $( \varphi^I , \chi^I)$ transform as a symplectic vector.
Then, $d_{\text{micro}}$ can be retrieved by the following inverse Laplace transform 
\begin{widetext}
 $$
  d_{\text{micro}}  ( P^I , Q_I , J ) = \oint \rd \varphi \? \rd \chi \? \rd \epsilon \?
  \delta \left( g_I \varphi^I - g^I \chi_I - \alpha \epsilon - 1 \right) 
  Z_{\text{can.}} ( \varphi^I , \chi_I , \epsilon ) \?
  e^{- \pi \ii \left( \varphi^I Q_I - \chi_I P^I - \ii \epsilon J \right)} \, ,
 $$
\end{widetext}
where $\rd \varphi \equiv \prod_{I} \rd \varphi^I$ and $\rd \chi \equiv \prod_{I} \rd \chi_I$.
Thus, in the limit of large charges, \cf\;\eqref{Me:entropy:function},
\be
 \label{Z:can}
 \log Z_{\text{can.}} ( \varphi^I , \chi_I , \epsilon ) \approx \frac{\pi \ii}{2} H ( \varphi^I , \chi_I , \epsilon ) \, ,
\ee
that is interpreted as the free energy of the black hole in the canonical ensemble.

To illustrate how the formalism works we will discuss several explicit examples in the following.
All these models admit an uplift on either $S^7$ to M--theory or on $S^5$ to type IIB string theory.
In the latter case $g_0 = 0$, and the AdS$_2 \times_w M_2$ near--horizon geometry arises upon considering a specific circle reduction (\ie\;a quotient of AdS$_3$ geometry).

{\it Twisted AdS$_4$ black holes \cite{Hristov:2018spe}}.---One class of BPS black holes are specified with \emph{non--zero} magnetic charge for the R--symmetry
and near--horizon AdS$_2 \times S^2_\epsilon$.
They are solutions of the $stu$ model with prepotential
\be
 \label{sqrt:F}
 F ( X^I ) = 2 \ii \sqrt{ X^0 X^1 X^2 X^3} \, ,
\ee
and a purely electric gauging, \ie\;$g_I \equiv g$ and $g^I = 0$ $(I = 0 , \ldots, 3)$.
We further set $g = 1$, hence fixing the AdS$_4$ scale $l_{\text{AdS}_4}^2 = \frac12$.
The quartic invariant $I_4$ for this model is given by \cite{Klemm:2012vm}
\bea
 \label{I4:sqrt}
 I_4 ( \Phi ) & = - \left( \varphi^0 \chi _0 - \varphi^1 \chi_1 - \varphi^2 \chi_2 - \varphi^3 \chi_3 \right)^2 \\
 & + 4 \big( \varphi^1 \varphi^2 \chi_1 \chi_2 + \varphi^1 \varphi^3 \chi_1 \chi_3 \\
 & + \varphi^2 \varphi^3 \chi_2 \chi_3 + \varphi^0 \varphi^1 \varphi^2 \varphi^3 + \chi _0 \chi _1 \chi _2 \chi _3 \big) \, .
\eea
For later convenience we define the following symplectic functions of the charges
\bea
 \label{F2:F3:Theta}
 F_2 & \equiv \sum_{I \neq J} \left( P^I P^J + 2 Q_I Q_J \right) - \sum_{I} ( P^I )^2 \, , \\
 F_3 & \equiv - \sum_{I} P^I Q_I \Big( 2 P^I - \sum_{I} P^I \Big) + \frac{1}{3} \sum_{\substack{I \neq J \neq K}} Q_I Q_J Q_K \, , \\
 \Theta & \equiv F_2^2 - 16 I_4 ( \Gamma )  \, .
\eea
The magnetic charges of the black hole satisfy the \emph{twisting condition} $\sum_{I} P^I = - 1$.
Moreover, to ensure regularity of the solution one must impose that $\sum_{I} Q_I = 0$ and $F_3 = 0$.
Thus, we are left with a six--parameter family of solutions, labeled for instance by $(P^0, P^1, P^2, Q_0 , Q_1, J )$.
The entropy is determined by the area law
\be
 \label{SBH:mAdS4}
 S_{\text{BH}} ( P^I , Q_I , J ) = \frac{\pi}{2 \sqrt{2}} \sqrt{F_2 + \sqrt{ \Theta - (4 J )^2}} \, .
\ee

For this class of black holes we have $\alpha = 0$, and therefore the constraint \eqref{const:phi:epsilon} simplifies to $\sum_{I } \varphi^I - 1 = 0$.
Let us denote by $( \mathring \Phi ,\mathring \epsilon )$ the critical points of the entropy functional \eqref{Me:entropy:function}.
Then, the attractor equations \eqref{PQJ:attractor} read
\footnote{We inherit from \cite{Hristov:2018spe} an unfavourable choice of sign for the prepotential \eqref{sqrt:F}
that leads to ambiguities when comparing with the literature.
To remedy this situation, we have chosen to work with the negative determination for the square root in $\cE_A (P^I , \varphi^I , \epsilon)$, see \eqref{hhz:A-gluing}.}
\be
 \label{attractor:A-twist}
 \Gamma = - \frac{1}{2 \mathring \epsilon} \re \left( \cX_{\text{NP}} - \cX_{\text{SP}} \right) , \qquad
 \mathring \Phi = - \frac{\ii}{2} \left( \cX_{\text{NP}} + \cX_{\text{SP}} \right) .
\ee
We verified that inserting $( \mathring \Phi ,\mathring \epsilon )$ into \eqref{Me:entropy:function} yields the entropy \eqref{SBH:mAdS4}.

{\it Kerr--Newman--AdS$_4$ black holes \cite{Cvetic:2005zi,Hristov:2019mqp}}.---Let us now consider the five-parameter family of BPS black holes with \emph{vanishing} magnetic charge for the R--symmetry and near--horizon AdS$_2 \times S^2_\epsilon$.
This model also belongs to the $stu$ model with the prepotential \eqref{sqrt:F} and a purely electric gauging.
The magnetic charges are subject to the constraints $P^0 = - P^1$ and $\sum_{I} P^I = 0$, which corresponds to the absence of a topological twist.
Moreover, the absence of closed time--like curves imposes the following non--linear constraint among the charges
\be
 \label{KN:PQJ:const}
 0 = 4 S_{\text{BH}}^4 - \pi^2 ( F_2 + 1 ) S_{\text{BH}}^2 + \pi^4 \left( I_4 ( \Gamma ) + J^2 \right) ,
\ee
where the entropy as a function of charges is given by
\be
 \label{SBH:KN}
 S_{\text{BH}} ( P^I , Q_I , J) = \frac{\pi}{\sqrt{2}} \sqrt{\frac{ F_3 + J }{\sum_{I = 0}^3 Q_I}} \, .
\ee

For this class of black holes we have $\alpha = \ii$, and thus the constraint \eqref{const:phi:epsilon} reads $\sum_{I} \varphi^I - \ii \epsilon - 1 = 0$,
that is a reflection of the non--linear constraint \eqref{KN:PQJ:const} among the conserved charges $(P^I , Q_I , J)$.
We find that the attractor equations \eqref{PQJ:attractor} are given by
\be
 \Gamma = \frac{1}{2 \mathring \epsilon} \re \left( \cX_{\text{NP}} - \wb \cX_{\text{SP}} \right) , \qquad
 \mathring \Phi = \frac{\ii}{2} \left( \cX_{\text{NP}} + \wb \cX_{\text{SP}} \right) ,
\ee
where $\wb \cX_{\text{SP}}$ denotes the complex conjugate of the symplectic vector of scalars evaluated at the South Pole of $S^2_\epsilon$.
Furthermore, we checked that $\cI ( \mathring \Phi , \mathring \epsilon )$ matches precisely the entropy \eqref{SBH:KN}.

{\it AdS$_2 \times_w \Sigma$ solutions in the $ F = - \ii X^0 X^1$ model}.---Let us now look at BPS black holes with near--horizon geometry AdS$_2 \times_w \Sigma$, where $\Sigma$ is a bad orbifold of $S^2$ called the \emph{spindle}.
They are solutions of the $\U( 1 )^2$ gauged $\cN = 2$ supergravity coupled to one vector multiplet,
with prepotential $F ( X^I ) = - \ii X^0 X^1$, and the purely electric gauging $G = \{ 0 , 0 ; 1 , 1 \}$.
The Hesse potential $I_4$ for this model is given by $\cH ( \Phi ) = - ( \varphi^0 \varphi^1 + \chi_0 \chi_1 )$ \cite{Klemm:2012yg}.
The solution has conical singularities with deficit angles $\frac{2 \pi}{n_1}$ and $\frac{2 \pi}{n_2}$ at the poles of $\Sigma$
\footnote{The angular momentum $J$ evaluated at the horizon depends crucially on the choice of pure gauge that can be added to the electric gauge fields.
We work in the gauge $A^I |_{\text{SP/NP}} = \mp P^I \rd \phi$ for $I=0,1$, where $\phi$ is the angular coordinate on the spindle with period $2 \pi$.}
$$
 \begin{aligned}
  p_{\text{SP}} & = -\frac{\left( 4 ( P^0 P^1 + Q_0 Q_1 ) + \frac{1}{2 n_1 n_2} \right) \left( 1 - \frac{2}{n_1} \frac{J}{F_3} \right)}{4 ( P^0 P^1 + Q_0 Q_1 ) + \frac{1}{n_1 n_2} } \, , \\
  p_{\text{NP}} & = \frac{\left( 4 ( P^0 P^1 + Q_0 Q_1 ) + \frac{1}{2 n_1 n_2} \right) \left( 1 - \frac{2}{n_2} \frac{J}{F_3} \right)}{4 ( P^0 P^1 + Q_0 Q_1 ) + \frac{1}{n_1 n_2} } \, ,
 \end{aligned}
$$
respectively, where $(n_1 , n_2 )$ are arbitrary coprime positive integers with $n_1 < n_2$.
$F_3$ is given in \eqref{F2:F3:Theta} when we set $P^3 = P^1$, $P^2 = P^0$, $Q_3 = Q_1$, and $Q_2 = Q_0$.
Magnetic charges satisfy the \emph{anti--twist} condition $P^0 + P^1 = \frac12 ( \frac{1}{n_1} - \frac{1}{n_2} )$,
therefore, there is a non--zero magnetic flux for the R--symmetry gauge field through the spindle \cite{Ferrero:2020twa}.
Moreover, to ensure regularity the conserved charges are subject to the following non--linear constraint
$$
 J = - \frac{1}{4} ( Q_0 + Q_1 ) \left( \chi - \sqrt{ 16 ( P^0 P^1 + Q_0 Q_1 ) + \chi^2} \right) ,
$$
where $\chi = \frac{1}{n_1} + \frac{1}{n_2}$ is the Euler number of the spindle $\Sigma$.
The entropy can be compactly written as
\be
 \label{SBH:Spindle:AdS4}
 S_{\text{BH}} ( P^I, Q_I , J ) = \pi \frac{J}{Q_0 + Q_1} \, .
\ee
For this class of black holes we set $\alpha = - \frac{\ii}{2} \chi$, and thus the constraint \eqref{const:phi:epsilon} reads $\sum_{I } \varphi^I + \frac{\ii}{2} \chi \epsilon - 1 = 0$.
We find that the attractor equations \eqref{PQJ:attractor} read
\bea
 \label{attractor:Sigma}
 \Gamma & = - \frac{1}{2 \mathring \epsilon} \re \left( \cX_{\text{NP}} \? p_{\text{NP}} + \wb \cX_{\text{SP}} \? p_{\text{SP}} \right) , \\
 \mathring \Phi & = - \frac{\ii}{2} \left( \cX_{\text{NP}} \? p_{\text{NP}} - \wb \cX_{\text{SP}} \? p_{\text{SP}} \right) .
\eea
We checked that inserting the above values into \eqref{Me:entropy:function} reproduces the entropy \eqref{SBH:Spindle:AdS4}.

{\it Twisted AdS$_5 \times S^5$ black strings \cite{Hosseini:2019lkt}}.---Let us consider the $stu$ model with prepotential
\be
 \label{cubic:F}
 F ( X^I ) = \frac{X^1 X^2 X^3}{X^0} \, ,
\ee
and a purely electric gauging, \ie\;$G = \{ {\bf 0} ; 0 , g \}$.
We also set $g = 1$, thus fixing the AdS$_5$ length scale $l_{\text{AdS}_{5}} = \sqrt{2}$ \footnote{We follow the conventions used in \cite[Sect.\,2.2]{Hosseini:2021fge}.}.
The quartic invariant $I_4$ is determined by \cite{Mohaupt:2011aa}
\bea
 \label{I4:cubic}
 I_4 ( \Phi ) & = - \left( \varphi^0 \chi_0 + \varphi^1 \chi_1 + \varphi^2 \chi_2 + \varphi^3 \chi_3 \right)^2 \\
 & + 4 \big( \varphi^1 \varphi^2 \varphi^3 \chi_0 + 4 \varphi^1 \varphi^2 \chi_1 \chi_2 \\
 & + 4 \varphi^1 \varphi^3 \chi_1 \chi_3 + 4 \varphi^2 \varphi^3 \chi_2 \chi_3 - 4 \varphi^0 \chi_1 \chi_2 \chi_3 \big) \, .
\eea
The $stu$ model described above may be viewed as the circle compactification of $\U(1)^3$ truncation of AdS$_5 \times S^5$.
There is a black string solution of the latter that is characterized by a \emph{non--zero} magnetic charge for the R--symmetry and near--horizon BTZ$\times S^2_\epsilon$.
Upon compactification on the BTZ circle, one obtains a four--dimensional black hole with angular momentum $J$,
and the symplectic vector of magnetic and electric charges $\Gamma = \{ 0 , P^i ; Q_I \}$ $(I = 0, \ldots, 3)$.
Here, $Q_0$ labels the momentum around the BTZ circle and, as is familiar in Kaluza--Klein (KK) reduction, a KK momentum becomes a conserved charge in lower dimension.
The magnetic charges are restricted to satisfy $\sum_i P^i = - 1$, and there is a further constraint among charges $\sum_{i} P^i Q_i \big( 2 P^i - \sum_{j} P^j \big) = 0$.
We are thus left with a six--parameter family of solutions, labeled for instance by $(P^0, P^1, Q_0 , Q_1, Q_2, J )$.
The Bekenstein--Hawking entropy as a function of charges is given by
$$
 S_{\text{BH}} ( P^i, Q_I , J ) =  \pi \sqrt{\frac{I_4( \Gamma ) + J^2}{ - \sum_{i} ( P^i )^2 + \sum_{\substack{i \neq j}} P^i P^j }} \, .
$$

We set $\alpha = 0$ in \eqref{const:phi:epsilon}, hence, the constraint among the chemical potentials reads $\sum_{i} \varphi^i - 1 = 0$.
We find that the attractor equations \eqref{attractor:A-twist} remain valid also in this case.
Moreover, we verified that $\cI ( \mathring \Phi , \mathring \epsilon ) \equiv S_{\text{BH}} ( P^i , Q_I , J )$.

{\it Black spindles in AdS$_5 \times S^5$ \cite{Ferrero:2020laf,Hosseini:2021fge,Boido:2021szx}}.---Amazingly enough, one can find near--horizon geometries BTZ$\times_w \Sigma \times S^5$ that are specified with the R--symmetry flux $\frac{1}{2 \pi} \int_\Sigma F^R = \frac12 ( \frac{1}{n_1} - \frac{1}{n_2} )$
through the spindle $\Sigma$.
This model also belongs to the $stu$ model with the prepotential \eqref{cubic:F} and a purely electric gauging.
The most general family of such black holes, with magnetic and electric charges $\Gamma = \{ 0 , P^i ; Q_I \}$ and angular momentum $J$, has not been written yet.
Two family of black holes, with either $Q_i = 0$ or $P^i = \frac{1}{6} (\frac{1}{n_1} - \frac{1}{n_2})$ for $i=1,2,3$, was found in \cite{Hosseini:2021fge}.
There is a constraint on the magnetic charges $\sum_{i} P^i = \frac{1}{2} (\frac{1}{n_1} - \frac{1}{n_2})$,
which corresponds to the fact that the theory is \emph{anti--twisted} along $\Sigma$, and
a further constraint involving the electric charges and angular momentum $\frac{\chi}{2} J = - \sum_{i} P^i Q_i \big( 2 P^i - \sum_{j} P^j \big)$,
that is reflected in the constraint \eqref{const:phi:epsilon} with $\alpha = - \frac{\ii}{2} \chi$.
Organizing the flavor symmetries in the basis $K_1 = Q_1 - Q_3$, $K_2 = Q_2 - Q_3$, and $K_3 = - J - \frac{1}{2} \chi Q_3$,
the entropy of the most general black hole is \emph{expected} to be given by the charged Cardy formula \cite{Hosseini:2020vgl}
\be
 \label{SBS:spindls}
 S_{\text{BH}} ( P^i , Q_I , J ) = \pi \sqrt{
 \frac{c^{\text{CFT}} (P^i)}{3 \sqrt{2}}
 \wt Q_0
 } \, ,
\ee
where $c^{\text{CFT}} (P^i)$ is the \emph{exact} central charge of the holographic dual $\cN = (0 , 2)$ CFT
$$
 c^{\text{CFT}} ( P^i ) = \frac{ 48 \sqrt{2} \? P^1 P^2 P^3}{\chi^2 - 4 \left( \sum_i ( P^i )^2 - \sum_{\substack{i \neq j}} P^i P^j \right)} \, .
$$
We also defined $\wt Q_0 \equiv Q_0 - \frac{1}{2 \sqrt{2}} \sum_{A , B} K_A ( k^{-1} )_{AB} K_B$, where $k_{AB}$ is the matrix of the levels of the abelian symmetries
$$
 k_{AB} =
 \left(
  \begin{array}{ccc}
   \sqrt{2} P^2 & \frac{P^1 + P^2 - P^3}{\sqrt{2}} & \frac{\chi }{2 \sqrt{2} } P^2 \\
   \frac{P^1 + P^2 - P^3}{\sqrt{2}} & \sqrt{2} P^1 & \frac{\chi }{2 \sqrt{2} } P^1 \\
    \frac{\chi }{2 \sqrt{2} } P^2 &  \frac{\chi }{2 \sqrt{2} } P^1 & - \sqrt{2} P^1 P^2 P^3 \\
  \end{array}
 \right) .
$$
We checked that, setting either $Q_i = 0$ or $P^i = \frac{1}{6} ( \frac{1}{n_1} - \frac{1}{n_2} )$ for $i=1,2,3$,
the attractor equations \eqref{attractor:Sigma} hold true
and the on--shell value of the entropy functional \eqref{Me:entropy:function} matches the entropy \eqref{SBS:spindls},
and are thus confident that the result is valid in general.

\begin{acknowledgements}
We thank Kiril Hristov and Alberto Zaffaroni for useful conversations and comments.
SMH is supported in part by the STFC Consolidated Grant ST/T000791/1.
\end{acknowledgements}

\bibliography{CovGrBlocks}

\end{document}